\documentclass[11pt]{article}

\usepackage{booktabs}
\usepackage{enumitem}
\usepackage{amsmath}
\usepackage{amssymb}
\usepackage{hyperref}
\usepackage{geometry}
\usepackage{graphicx}
\usepackage{tikz}
\usepackage{microtype}

\microtypesetup{expansion=false}
\newcommand{\Description}[1]{}

\title{Human-Centered AI for Safe Shuttle Car Routing in Underground Room-and-Pillar Coal Mines Using Graph Neural Networks}
\author{Bryant Pollard\\Clemson University\\\texttt{bdpolla@clemson.edu}}
\date{April 2026}

\begin{document}
\maketitle

\begin{abstract}
Underground room-and-pillar coal mining requires shuttle car operators to make safety-critical routing decisions under conditions of low visibility, dynamic miner movement, congestion, and limited real-time information. This paper presents a human-centered AI decision-support system that recommends safe shuttle car routes using a Graph Neural Network (GNN) trained on expert-informed synthetic data and deployed through a browser-based interface backed by cloud inference services. Rather than making a purely model-centric contribution, the paper shows how interviews, participatory design, usability testing, interaction logs, and interpretability analysis shaped both the interface and the AI model. The resulting application evolved beyond route recommendation to include inline user feedback, blockage reporting, audio cues, and SHAP-based interpretability within a single interactive system. Evaluation across two usability sessions with six participants showed improved task completion, faster response times, fewer errors, higher usability scores, lower cognitive load, and stronger agreement with AI recommendations in the later version. The paper contributes a focused example of how human-centered design can transform an AI routing prototype into a more transparent, auditable, and safety-supportive decision-support system.
\end{abstract}

\section{Introduction}
Underground coal mining is a safety-critical environment in which operators routinely make navigation decisions under uncertainty. Shuttle car routing is especially challenging because drivers must coordinate with the miner, other shuttle cars, and changing mine conditions while working with poor visibility and incomplete situational information. In this setting, a useful AI system should not replace operator judgment. Instead, it should provide decision support that improves situational awareness, surfaces safer path recommendations, and preserves human agency.

This paper presents the design and evaluation of a human-centered AI system for shuttle car route recommendation in a room-and-pillar coal mine setting. The system uses a Graph Neural Network (GNN) to recommend safe routes on a graph representation of the mine and delivers those recommendations through a browser-based interface connected to cloud-hosted inference services. The paper focuses on one central contribution: showing how human-centered AI methods transformed a route-recommendation prototype into a more usable and transparent decision-support system. Interviews and participatory design clarified the operational problem and mine logic. Usability testing then exposed weaknesses in feedback capture, blockage handling, and notification design. Those findings directly shaped the deployed system: the later version added in-context feedback capture, multiple-blockage reporting, audio navigation cues, and SHAP-based interpretability support to improve transparency and user trust.

The central claim of the paper is that the value of the system lies not only in the predictive performance of the GNN, but also in how the AI model, interface, and evaluation methods were integrated into a coherent human-centered decision-support workflow. Accordingly, the paper emphasizes the design decisions that shaped the final application, the system components required to support it, and the measured effects of those revisions.

The paper addresses three research questions. \textbf{RQ1:} How can human-centered design methods inform the development of an AI route recommendation system for underground shuttle car operation? \textbf{RQ2:} How does the resulting system change after iterative feedback from interviews, participatory design, and usability testing? \textbf{RQ3:} To what extent does the final system provide evidence of usability, safety support, and interpretable AI behavior in a controlled evaluation setting?

The paper makes three contributions. First, it presents a human-centered AI prototype that integrates graph-based route recommendation, browser-based interaction, and model interpretability. Second, it documents how the system changed through human-centered methods. Third, it reports a focused mixed-method evaluation that combines usability metrics, safety perception, interaction logs, and model-validation evidence. The remainder of the paper is structured as follows: Related Work positions the contribution in mining, HCI, and HCAI literature; the next sections describe the problem setting, system, methodology, and model; Results and Discussion then report and interpret the empirical findings.

\section{Related Work}
This work sits at the intersection of mining safety, intelligent navigation support, and human-centered AI. At the mining-domain level, prior work on room-and-pillar operations has described the structural and operational constraints that shape mine layout, ventilation, and traversability \cite{kobylkin2020design}. That line of work is referenced here to justify why route reasoning in this setting must be layout-aware rather than treated as a generic navigation problem. At the task level, prior research on underground shuttle car navigation has explored autonomous or semi-autonomous routing support \cite{androulakis2021development}. That work is relevant because it establishes shuttle car routing as an appropriate AI application domain, even though the present paper frames the problem as decision support rather than full autonomy.

From an HCI and HCAI perspective, the paper is grounded in scholarship arguing that AI systems should be reliable, safe, and centered on human needs rather than model performance alone \cite{shneiderman2020hcai}. This citation justifies the overall framing of the project as a human-centered AI system rather than a model-only contribution. The paper also draws on human-AI interaction guidance emphasizing clear system communication, user control, feedback loops, and support for breakdown recovery during use \cite{amershi2019guidelines}. That guidance is especially relevant here because the interface must help operators understand recommendations, provide feedback, and recover from deviations under safety-critical conditions. Together, these perspectives support the paper's emphasis on preserving operator authority and encouraging users to compare their own judgment with AI recommendations rather than relying on the system uncritically.

The interpretability component is grounded in prior work on additive feature-attribution methods for interpreting model predictions \cite{lundberg2017unified}. In this paper, that citation supports the choice to use SHAP as one part of a broader human-centered workflow that also includes interface design, usability evaluation, and interaction logging, rather than treating explanation as a sufficient solution by itself.

Taken together, these literatures suggest an important gap. Mining and navigation research establishes the operational difficulty of underground routing and motivates AI assistance, but it tends to emphasize technical navigation capability. HCAI and human-AI interaction research, by contrast, emphasizes trust calibration, transparency, user control, and workflow fit, but usually does not address underground mining as a deployment setting. The present paper contributes at that intersection by treating shuttle car routing as a socio-technical design problem: route quality matters, but so do explanation, feedback capture, state awareness, and the operator's ability to accept, question, or override a recommendation. 

The present work differs from a purely autonomous navigation framing in two ways. First, it treats route recommendation as a decision-support problem in which the human operator remains in the loop as the final decision maker, rather than as a fully automated navigation system. Second, it combines model development with interface design, interpretability analysis, and usability evaluation. That shift matters in a safety-critical setting because the practical usefulness of the system depends on trust calibration, interpretability, ease of use, and fit with existing work practices as much as on route prediction accuracy.

\section{Problem Setting and System}
\subsection{Mine Representation as a Graph}
The mine unit is represented as a grid-like graph in which nodes correspond to traversable positions and edges correspond to allowed shuttle car movement. This structure makes the problem naturally compatible with graph-based learning. The graph representation also reflects the underlying logic of the mine: a feeder location, cross cuts, faces, and a constrained set of safe routes between the feeder and the miner. These safe routes are specific to each shuttle car and change as the miner progresses across the face.

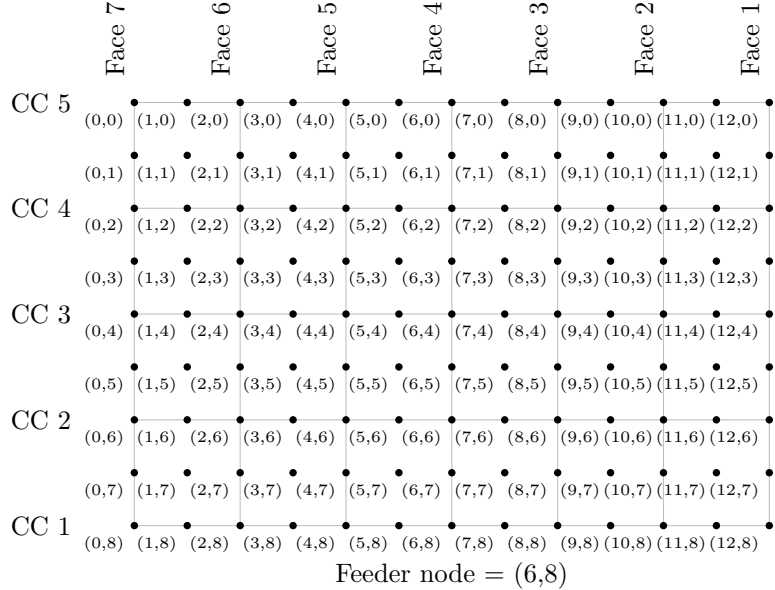
\begin{figure}[ht]
\centering
\begin{tikzpicture}[scale=0.7]
    \foreach \x in {0,...,12} {
        \foreach \y in {0,...,8} {
            \pgfmathtruncatemacro{\ylab}{8-\y}
            \pgfmathtruncatemacro{\xodd}{mod(\x,2)}
            \pgfmathtruncatemacro{\ylabodd}{mod(\ylab,2)}
            \ifnum\x<12
                \pgfmathtruncatemacro{\xnextodd}{mod(\x+1,2)}
                \pgfmathtruncatemacro{\ylabnext}{8-\y}
                \pgfmathtruncatemacro{\ylabnextodd}{mod(\ylabnext,2)}
                \ifnum\xodd=1
                    \ifnum\ylabodd=1
                    \else
                        \ifnum\xnextodd=1
                            \ifnum\ylabnextodd=1
                            \else
                                \draw[gray!50] (\x,\y) -- (\x+1,\y);
                            \fi
                        \else
                            \draw[gray!50] (\x,\y) -- (\x+1,\y);
                        \fi
                    \fi
                \else
                    \ifnum\xnextodd=1
                        \ifnum\ylabnextodd=1
                        \else
                            \draw[gray!50] (\x,\y) -- (\x+1,\y);
                        \fi
                    \else
                        \draw[gray!50] (\x,\y) -- (\x+1,\y);
                    \fi
                \fi
            \fi
            \ifnum\y<8
                \pgfmathtruncatemacro{\ylabbelow}{8-(\y+1)}
                \pgfmathtruncatemacro{\ylabbelowodd}{mod(\ylabbelow,2)}
                \ifnum\xodd=1
                    \ifnum\ylabodd=1
                    \else
                        \ifnum\xodd=1
                            \ifnum\ylabbelowodd=1
                            \else
                                \draw[gray!50] (\x,\y) -- (\x,\y+1);
                            \fi
                        \else
                            \draw[gray!50] (\x,\y) -- (\x,\y+1);
                        \fi
                    \fi
                \else
                    \ifnum\xodd=1
                        \ifnum\ylabbelowodd=1
                        \else
                            \draw[gray!50] (\x,\y) -- (\x,\y+1);
                        \fi
                    \else
                        \draw[gray!50] (\x,\y) -- (\x,\y+1);
                    \fi
                \fi
            \fi
            \fill (\x,\y) circle [radius=0.07];
            \node[font=\tiny, anchor=north east] at (\x,\y) {(\x,\ylab)};
        }
    }
    \node[font=\small, anchor=north] at (6,-0.5) {Feeder node = (6,8)};
    \node[anchor=east, font=\small] at (-1,0) {CC 1};
    \node[anchor=east, font=\small] at (-1,2) {CC 2};
    \node[anchor=east, font=\small] at (-1,4) {CC 3};
    \node[anchor=east, font=\small] at (-1,6) {CC 4};
    \node[anchor=east, font=\small] at (-1,8) {CC 5};
    \node[anchor=south, font=\small, rotate=90] at (0,9.2) {Face 7};
    \node[anchor=south, font=\small, rotate=90] at (2,9.2) {Face 6};
    \node[anchor=south, font=\small, rotate=90] at (4,9.2) {Face 5};
    \node[anchor=south, font=\small, rotate=90] at (6,9.2) {Face 4};
    \node[anchor=south, font=\small, rotate=90] at (8,9.2) {Face 3};
    \node[anchor=south, font=\small, rotate=90] at (10,9.2) {Face 2};
    \node[anchor=south, font=\small, rotate=90] at (12,9.2) {Face 1};
\end{tikzpicture}
\caption{Mine unit represented as a graph with labeled nodes, cross cut rows, and face columns. Edges represent traversable shuttle car paths.}
\label{fig:mine-graph}
\end{figure}

Figure~\ref{fig:mine-graph} shows the graph representation used for route reasoning.

\subsection{System Overview}
The deployed system consists of four tightly coupled parts. First, a browser-based interface presents the mine state, route recommendations, and interaction controls. Second, a GNN model generates shuttle-car-specific path recommendations. Third, an AWS-based deployment pipeline connects the interface to inference and logging services. Fourth, an evaluation workflow collects both quantitative and qualitative evidence about system use.

The system was designed as decision support rather than automation. The operator remains responsible for the final route decision, while the system provides route guidance, state visibility, feedback capture, and safety-oriented alerts.

\subsection{Human-Centered Design Objectives}
Three design objectives guided the final system. The first was \textit{human agency}: the system should recommend routes without taking control away from the operator. The second was \textit{transparency}: users should be able to see what the system is recommending, respond to it, and review logged behavior. The third was \textit{safety support}: the system should prioritize safer operation, miner movement, and blockage conditions rather than optimize only for speed, automation, or the most profitable outcome.

\section{Methodology}
\subsection{Human-Centered Design Process}
The project used a staged human-centered design process. Semi-structured interviews with miners and shuttle car operators established the initial problem framing, safety concerns, route logic, and operational constraints. Participatory design sessions were then used to validate the mine layout, interface assumptions, and route behaviors. Finally, two usability testing sessions with six participants were conducted using the deployed system. Supporting evidence for these activities is provided in Appendix~\ref{app:interviews}, Appendix~\ref{app:design}, Appendix~\ref{app:usability}, and Appendix~\ref{app:usability-trial2}.

The later version of the system changed directly because of this process. Feedback led to a more usable feedback loop, multiple-blockage support, audibly reinforced alerts, and a more interpretable AI model.

\subsection{Participants and Data Collection}
The evaluation involved domain experts during early design activities and six repeated participants across two usability sessions. To mitigate bias in the knowledge used to shape the system, the study sought driver perspectives from participants of different genders and from different geographic mining regions. Each session was conducted as a controlled simulation of shuttle car operation across the mine unit. Participants completed navigation tasks in the browser interface while route decisions, deviations, timing, and interaction events were logged automatically. Post-session data included System Usability Scale (SUS) responses, cognitive load ratings, safety perception ratings, and open-ended feedback. The questionnaires and detailed participant materials are reproduced in Appendix~\ref{sec:questionnaires}.

\subsection{Evaluation Measures}
The quantitative measures used during usability testing were tasks completed, response time per task, errors, agreement rate, adjusted agreement rate, SUS score, cognitive load, and safety perception. Qualitative evidence included open-ended comments and interface-collected feedback. Agreement Rate is the percentage of tasks in which the tester followed the GNN model's recommended route without deviation. Adjusted Agreement Rate excludes corrective steps and confirmed mistakes needed to return to the safe route.

These measures were selected to reflect both system performance and human-centered use quality. Task completion, response time, and error counts capture operational performance. Agreement measures capture how often users followed the AI recommendation and whether safe-route completion was preserved even when deviations occurred. SUS, cognitive load, and safety perception provide complementary evidence about whether the system was understandable, manageable to use, and perceived as safety-supportive.

\section{Model and Deployment}
\subsection{Synthetic Data Generation}
Because real operational route data were not readily available in a form suitable for training, the project used synthetic route data derived from expert operational knowledge. This allowed domain-specific routing preferences, shuttle-car-specific behavior, and miner-position changes to be encoded in a supervised learning model while staying within the practical constraints of the project, which was limited to a seven-face mine unit with three shuttle cars (A, B, and C).

An example of the synthetic training data is shown below.

\begin{verbatim}
shuttle_car,start_x,start_y,dest_x,dest_y,scenario,path_node_ids,path_coords
A,6,8,12,3,dest_12_3,110 111 112 99 86 87 88 75 62 63 64 51,
"(6,8) (7,8) (8,8) (8,7) (8,6) (9,6) (10,6) (10,5) (10,4) (11,4) (12,4) (12,3)"
\end{verbatim}

The resulting dataset supported 16,800 synthetic training samples across seven destinations and three shuttle cars.

\subsection{Graph Neural Network Architecture}
The route recommendation model is a GNN tailored to the graph structure of the mine unit. Each node receives a 9-dimensional feature vector containing positional and car-specific information. The model uses six graph convolutional layers with 512 hidden units and ReLU activation, followed by a linear output layer that predicts the next recommended node. Ultra-light dropout of 0.05 is applied after each graph convolutional layer, and training uses cross-entropy loss with the Adam optimizer and a learning rate of $5\times10^{-4}$.

The trained model learned distinct shuttle-car-specific strategies: Car A generally used routes to the right of the feeder, Car B used more direct routes, and Car C used safer routes to the left of the feeder. Example route predictions are shown in Table~\ref{tab:route-validation}.

\begin{table}[ht]
\centering
\caption{Sample route predictions for selected miner destinations and shuttle cars}
\label{tab:route-validation}
\small
\setlength{\tabcolsep}{4pt}
\begin{tabular}{p{0.10\linewidth} p{0.08\linewidth} p{0.08\linewidth} p{0.66\linewidth}}
\toprule
\textbf{Miner} & \textbf{Car} & \textbf{Steps} & \textbf{Predicted Route} \\
\midrule
(12,3) & A & 12 & (6,8)$\rightarrow$(7,8)$\rightarrow$(8,8)$\rightarrow$(8,7)$\rightarrow$(8,6)$\rightarrow$(9,6)$\rightarrow$(10,6)$\rightarrow$(10,5)$\rightarrow$(10,4)$\rightarrow$(11,4) $\rightarrow$(12,4)$\rightarrow$(12,3) \\
       & B & 12 & (6,8)$\rightarrow$(6,7)$\rightarrow$(6,6)$\rightarrow$(7,6)$\rightarrow$(8,6)$\rightarrow$(8,5)$\rightarrow$(8,4)$\rightarrow$(9,4)$\rightarrow$(10,4)$\rightarrow$(11,4) $\rightarrow$(12,4)$\rightarrow$(12,3) \\
(8,3)  & A & 8 & (6,8)$\rightarrow$(7,8)$\rightarrow$(8,8)$\rightarrow$(8,7)$\rightarrow$(8,6)$\rightarrow$(8,5)$\rightarrow$(8,4)$\rightarrow$(8,3) \\
       & B & 8 & (6,8)$\rightarrow$(6,7)$\rightarrow$(6,6)$\rightarrow$(7,6)$\rightarrow$(8,6)$\rightarrow$(8,5)$\rightarrow$(8,4)$\rightarrow$(8,3) \\
(6,3)  & B & 6 & (6,8)$\rightarrow$(6,7)$\rightarrow$(6,6)$\rightarrow$(6,5)$\rightarrow$(6,4)$\rightarrow$(6,3) \\
       & C & 10 & (6,8)$\rightarrow$(5,8)$\rightarrow$(4,8)$\rightarrow$(4,7)$\rightarrow$(4,6)$\rightarrow$(4,5)$\rightarrow$(4,4)$\rightarrow$(5,4)$\rightarrow$(6,4)$\rightarrow$(6,3) \\
\bottomrule
\end{tabular}
\end{table}

\subsection{Interpretability and Deployment}
SHAP was used to analyze feature importance after model validation. As reported in Appendix~\ref{sec:shap-analysis}, the most important features for the first routing decision were destination x-position and shuttle identity, which is consistent with the intended problem structure. This interpretability step was important because it provided evidence that the model was using routing-relevant factors rather than spurious ones. In addition, the interface includes an information icon in the navigation suggestions so users can inspect the SHAP-based features associated with each decision.

The deployed system uses AWS services to deliver live recommendations to the browser interface. The architecture includes cloud storage and delivery for the front end, Lambda-based orchestration, API connectivity, and a SageMaker inference endpoint. 

\begin{figure}[ht]
\centering
\includegraphics[width=0.92\textwidth]{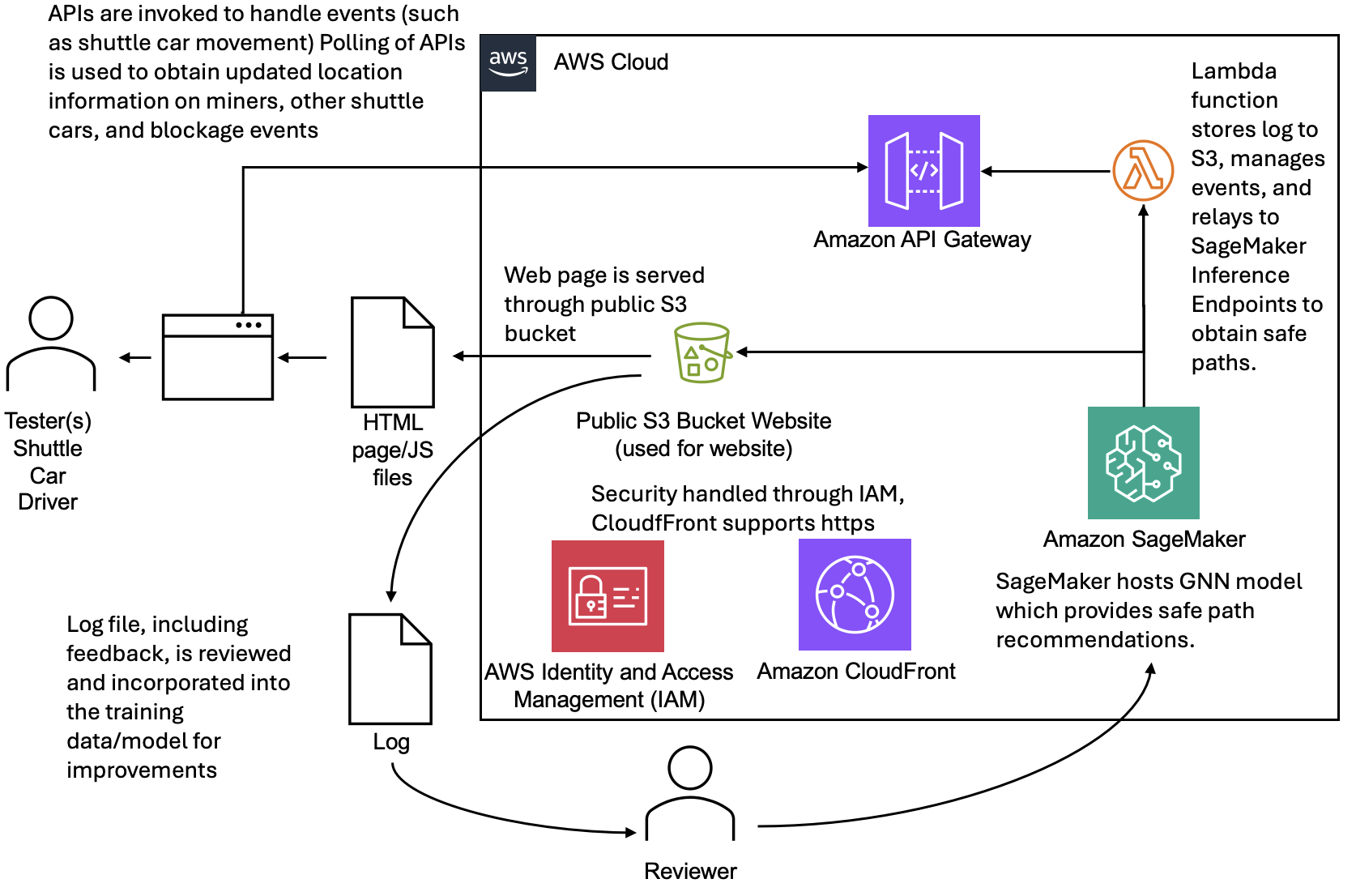}
\caption{Deployed architecture for the human-centered AI shuttle car routing system. The architecture connects the browser interface, AWS services, and the GNN inference endpoint for real-time route recommendation.}
\label{fig:hcai-architecture}
\end{figure} 

\section{Results}
\subsection{Design Revisions from Human Feedback}
Early interviews and participatory design identified route-safety concerns related to dynamic miner positions, limited visibility, traffic behavior, and the possibility of multiple blockages. Usability testing then documented additional issues in the interaction design. Table~\ref{tab:feedback-themes} summarizes the resulting revisions that were incorporated into the evaluated system.

\begin{table}[ht]
\centering
\caption{Human feedback themes and their effect on the final system}
\label{tab:feedback-themes}
\begin{tabular}{p{0.33\linewidth} p{0.60\linewidth}}
\toprule
\textbf{Feedback Theme} & \textbf{Impact on the System} \\
\midrule
\addlinespace[0.5em]
System needed to be more robust & Several issues were addressed, including boundary problems, unexpected behavior that required restarting, and lack of support for the secure HTTPS protocol. \\
\addlinespace[0.5em]
Feedback loop was difficult to use & Feedback was redesigned so that users could provide input while navigation was occurring instead of relying on a separate after-the-fact process. \\
\addlinespace[0.5em]
Multiple blockages can occur, and shuttle cars cannot navigate through them & The interface was expanded from supporting a single blockage to supporting multiple blockages marked as restricted areas. \\
\addlinespace[0.5em]
Notifications could be improved & Audible notifications were added to complement visual and textual cues and improve real-time situational awareness. \\
\bottomrule
\end{tabular}
\end{table}

\begin{figure}[ht]
\centering
\includegraphics[width=0.88\textwidth]{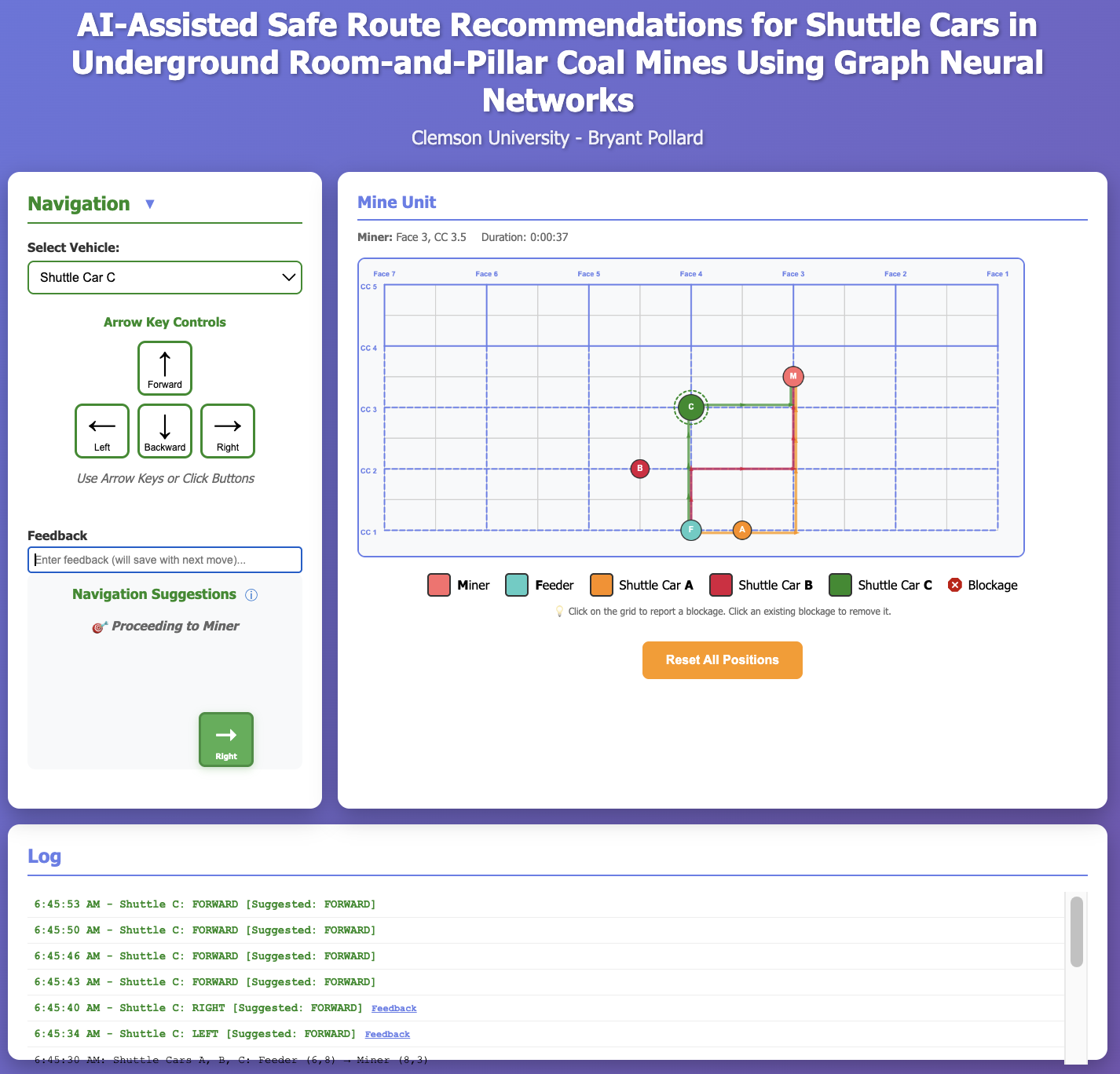}
\caption{Final interface used for evaluation. The interface supports route guidance, in-context feedback capture, customizable views, and audio cues for navigation events.}
\label{fig:ui-checkpoint3-main}
\end{figure}

Figure~\ref{fig:ui-checkpoint3-main} shows the evaluated interface after those revisions were incorporated.

\subsection{Usability Outcomes}
Two usability sessions were conducted with the same six participants. Table~\ref{tab:trial-improvements} reports the descriptive change between the earlier and later evaluated system versions.

\begin{table}[ht]
\centering
\caption{Consolidated improvement from Session 1 to Session 2 based on appendix usability tables}
\label{tab:trial-improvements}
\begin{tabular}{p{0.28\linewidth} p{0.16\linewidth} p{0.16\linewidth} p{0.16\linewidth} p{0.18\linewidth}}
\toprule
\textbf{Measure} & \textbf{Session 1} & \textbf{Session 2} & \textbf{Change} & \textbf{Observed Difference} \\
\midrule
Tasks completed & 13.8/14 & 14.0/14 & +0.2 tasks & Session 2 average was higher by 0.2 tasks. \\
Response time per task & 37.3 s & 32.8 s & -4.5 s & Session 2 average was lower by 4.5 seconds. \\
Errors & 0.67 & 0.17 & -0.50 & Session 2 average was lower by 0.50 errors. \\
Agreement rate & 91.8\% & 99.3\% & +7.5 pts & Session 2 average was higher by 7.5 percentage points. \\
Adjusted agreement rate & 100\% & 100\% & 0.0 pts & Both sessions had the same average. \\
SUS score & 82.5 & 88.8 & +6.3 pts & Session 2 average was higher by 6.3 points. \\
Cognitive load & 3.0/7 & 2.0/7 & -1.0 & Session 2 average was lower by 1.0 point. \\
Safety perception & 4.3/5 & 4.7/5 & +0.4 pts & Session 2 average was higher by 0.4 points. \\
\bottomrule
\end{tabular}

\vspace{0.5em}
\footnotesize
\textit{Note:} Agreement Rate is the percentage of tasks in which the tester followed the GNN model's recommended route without deviation. Adjusted Agreement Rate excludes corrective steps and confirmed mistakes needed to return to the safe route.
\end{table}

\subsection{Model Validation}
Table~\ref{tab:training-convergence} reports training convergence across the evaluated synthetic miner-car combinations. By epoch 3000, the model reached exact path matching across all 21 combinations in the reported setting.

\begin{table}[ht]
\centering
\caption{Training convergence of the enhanced GNN model}
\label{tab:training-convergence}
\begin{tabular}{lccc}
\toprule
\textbf{Epoch} & \textbf{Perfect Combinations} & \textbf{Min Accuracy} & \textbf{Average Loss} \\
\midrule
1000 & 5/21 & 0.400 & 1.5265 \\
2000 & 7/21 & 0.286 & 0.3911 \\
3000 & 21/21 & 1.000 & 0.0002 \\
\bottomrule
\end{tabular}
\end{table}

Table~\ref{tab:ai-evidence} summarizes the main forms of AI validation evidence reported in this study.

\begin{table}[ht]
\centering
\caption{Validation evidence for the AI solution}
\label{tab:ai-evidence}
\begin{tabular}{p{0.31\linewidth} p{0.60\linewidth}}
\toprule
\textbf{Evidence Type} & \textbf{What It Shows} \\
\midrule
Training dataset construction & Expert-informed synthetic scenarios were translated into structured route examples suitable for supervised learning. \\
Training progress summary & The model reached 100\% exact path matching across all 21 shuttle-car and destination combinations by epoch 3000 in the reported training configuration. \\
Route prediction examples & The system generated miner-car-specific safe paths rather than a single universal route. \\
Application integration & The trained model was deployed through an AWS SageMaker inference endpoint and connected to the browser-based interface for live recommendations. \\
Human-in-the-loop testing & Users generally followed the model's recommendations, and evaluation metrics suggest strong usability and safety perception. \\
Interpretability analysis & SHAP analysis showed that the model's first routing decision depended mainly on destination x-position and shuttle identity, which is consistent with domain expectations. \\
\bottomrule
\end{tabular}
\end{table}

\section{Discussion}
\subsection{Impact of Human-Centered AI}
The final system is materially different from the earlier model-driven prototype because of human-centered AI methods. The early version primarily demonstrated that route recommendations could be generated and delivered. The revised version is a broader decision-support system shaped by human agency, transparency, and safety. Human feedback changed what the system does, how it presents information, and how it is evaluated.

Several feedback themes shaped the revised system. First, the application became more robust by addressing stability issues and enforcing restrictions in blocked areas. Second, the feedback feature was redesigned so that users could provide feedback during navigation rather than only after completing a route. Third, the system moved from limited event handling to more realistic operational support through multiple-blockage reporting. Fourth, SHAP made the AI model more transparent by showing that the model relied on the intended routing parameters and by exposing that information in the interface.

These changes matter because the goal in this setting is not full autonomy. The goal is to support operators in making safer decisions while respecting their expertise. That design stance is the paper's main human-centered AI contribution.

The results provide tentative answers to the research questions introduced earlier. For \textbf{RQ1}, the design process shows that interviews and participatory design were effective for eliciting route logic, safety concerns, and interface needs that would have been difficult to derive from model development alone. For \textbf{RQ2}, the comparison across usability sessions shows that the system changed in practically meaningful ways after user feedback, particularly in the feedback loop, blockage handling, and audio cues. For \textbf{RQ3}, the final evaluation provides encouraging evidence that the system was usable, perceived as safety-supportive, and aligned with intended routing logic, although that evidence remains limited to a controlled synthetic setting.

\subsection{Limitations}
The main limitation is that the model was trained on synthetic data rather than real operational logs. This was a practical decision, but it limits claims about generalization to real-world mine conditions. The mine layout is also fixed to a seven-face unit, which simplifies the routing problem relative to real operations. In addition, the model exhibits overfitting to the encoded safe routes, and it does not yet generalize well to unseen path configurations such as complex blockage-driven rerouting. The deployment is appropriate for controlled demonstration and evaluation use, but a real underground deployment would require substantial additional work in sensing, networking, ruggedized hardware, and field validation.

\subsection{Threats to Validity}
Several threats to validity should be considered when interpreting the results. First, the usability evaluation used a small repeated sample of six participants, which is reasonable for formative HCI testing but limits statistical power and external validity. Second, because the same participants returned for the later session, some of the observed improvement may reflect familiarity with the interface in addition to design improvements. Third, the evaluation was conducted in a controlled simulation rather than in an active mine, so the reported outcomes do not capture the full variability of underground operations, communication interruptions, or sensing uncertainty. Fourth, the AI validation is based on synthetic route data and selected route examples, which support proof-of-concept claims but not broad real-world performance claims. These threats do not negate the value of the findings, but they do mean the results should be interpreted as evidence of feasibility and human-centered improvement rather than definitive field-level effectiveness.

\subsection{Future Work}
Future work should focus on incorporating real operational data, improving generalization to unseen routing conditions, and expanding the mine representation to support more flexible layouts and evolving conditions. It would also be valuable to investigate more advanced human-AI teaming designs, including improved explanations, better support for exception handling, and hardware integration for real-world deployment.

\section{Conclusion}
This paper presented a human-centered AI system for shuttle car route recommendation in underground room-and-pillar coal mining. The work combined a graph-based route-recommendation model, an interactive browser interface, cloud deployment, human-centered design methods, and interpretability analysis in a single evaluated application. The main contribution is not only that a GNN can learn shuttle-car-specific routes in a structured mine representation, but that the system changed meaningfully as a result of human-centered AI methods. Interviews, participatory design, usability testing, and interpretability analysis transformed the prototype into a more transparent, auditable, and safety-supportive decision-support system. The paper therefore offers a focused example of how a short-form HCAI study can connect system design, model transparency, and formative user evaluation in a safety-critical domain.

\section*{Acknowledgments}
Special thanks to my father, Dale Pollard, for his support and participation. I also thank Carlos Toxtli Hernandez, PhD, of Clemson University for teaching the course in which this project was developed and for assigning the project that motivated this work. I also sincerely thank everyone who participated in the semi-structured interviews, participatory design activities, and usability testing sessions.

\newpage
\appendix

\section{Semi-Structured Interview Guide and Feedback}
\label{app:interviews}
\subsection*{Interview Participant 1}
\textbf{Background:} Male with 12 years as a shuttle car operator and more than 20 years of experience as a miner; works the day shift in a room-and-pillar mining unit in Kentucky.\\
\textbf{Key Feedback:}
\begin{itemize}
    \item "Visibility is a big challenge. It is difficult to see from the driver's seat due to its low (profile) position. There are (plastic) curtains (that direct airflow), dust, and poor lighting obstructing the view."
    \item "I rely on radio calls and miner position to pick my route, but traffic jams happen when someone takes a shortcut."
    \item "Providing the locations of other shuttle cars and miners would be extremely helpful. The paths between the feeder and miner are known, but occasionally someone takes an unexpected route....This occurs when someone is sleepy or distracted."
    \item "Shuttle car paths overlap and are still considered safe. Drivers wait for the other car to leave before proceeding."
\end{itemize}
\begin{figure}[ht]
    \centering
    \includegraphics[width=0.7\textwidth]{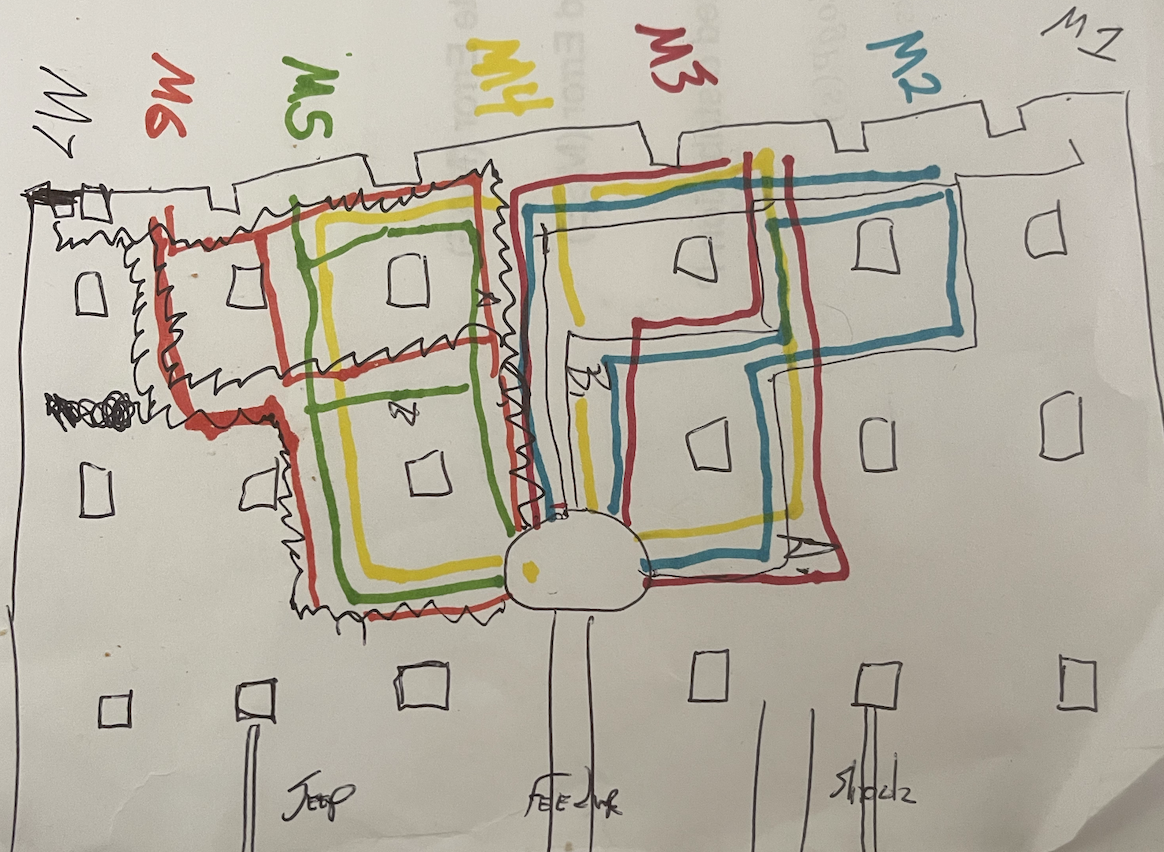}
    \caption{Mine unit layout sketch obtained during the semi-structured interview with Participant 1. This image was confirmed in the participatory design meeting with other experts.}
    \Description{A hand-drawn sketch of the mine unit layout showing the feeder, shuttle car routes, and the face structure discussed during the interview.}
    \label{fig:mineunit}
\end{figure}

\subsection*{Interview Participant 2}
\textbf{Background:} Male with 5 years as a miner operator who is also trained as a shuttle car operator in Kentucky.\\
\textbf{Key Feedback:}
\begin{itemize}
    \item "I follow the same path every time unless there's an issue or someone tells me to change."
    \item "The safest path is determined by the miner position. As the miner moves across the face, the paths change."
    \item "I'd trust AI if it explained why a route is safer, especially if it avoids congestion."
    \item "Most dangerous spots are near the belt line and the miner where visibility is lowest and congestion is high."
\end{itemize}

\section{Participatory Design Guide and Feedback}
\label{app:design}
\subsection*{Participant 1}
\textbf{Role:} Female shuttle car operator with 2 years of experience in Kentucky.\\
\textbf{Feedback:}
\begin{itemize}
    \item "All shuttle car drivers (male and female) follow the same routes and procedures."
    \item "The routes presented matched the routes I would have taken."
\end{itemize}
\subsection*{Participant 2}
\textbf{Role:} Male mechanic with 15 years of experience in Indiana.\\
\textbf{Key Feedback:} "Real-time equipment tracking would be helpful. Often miners use mine cap lights to signal other machine operators."

\subsection*{Participant 3}
\textbf{Role:} Male shuttle car operator with 30 years of mining experience in Kentucky.\\
\textbf{Key Feedback:} "There's little need for AI. Drivers know the routes and procedures well."

\subsection*{Participant 4}
\textbf{Role:} Male face boss with 10 years of experience in Kentucky.\\
\textbf{Key Feedback:} "In addition to monitoring shuttle car operations from the cars, it would be ideal if this was visible from the shack (or common area)."

\section{Usability Testing Guide and Results - Session 1}
\label{app:usability}
\subsection*{Tester 1 (T1)}
\textbf{Role:} Female shuttle car operator with 2 years of experience in Kentucky.\\
\textbf{Scenario Results:} Completed one full pass of mining across the face in shuttle car A and only accidentally deviated from the planned route.\\
\textbf{Key Feedback:} "The recommended paths and navigation suggestions were helpful and accurate."

\subsection*{Tester 2 (T2)}
\textbf{Role:} Male mechanic with 15 years of experience in Indiana.\\
\textbf{Scenario Results:} "I was off the suggested path once because the miner began moving while I was returning to the feeder with a load."\\
\textbf{Key Feedback:} "Sometimes the route wasn't obvious because the paths were on top of each other."

\subsection*{Tester 3 (T3)}
\textbf{Role:} Male face boss with 10 years of experience in Kentucky.\\
\textbf{Scenario Results:} Completed all tasks, but hesitated on congested routes.\\
\textbf{Key Feedback:} "The system was helpful in avoiding other shuttle cars. Knowing the location of other machinery would also be helpful."

\subsection*{Tester 4 (T4)}
\textbf{Role:} Male Shuttle car operator, 12 years of experience in Kentucky.\\
\textbf{Scenario Results:} Completed all tasks and found the interface easy to use.\\
\textbf{Key Feedback:} "The AI's route suggestions matched the routes I would have taken."

\subsection*{Tester 5 (T5)}
\textbf{Role:} Male Miner operator, 5 years of experience in Kentucky.\\
\textbf{Scenario Results:} Completed most tasks but struggled with map orientation.\\
\textbf{Key Feedback:} "The controls for the browser make sense, but are very different than an actual shuttle car. In my shuttle car, I flip positions and always move forward, so the concept of going backwards is relative to the position of the driver."

\subsection*{Tester 6 (T6)}
\textbf{Role:} Male technical user with previous mining experience in Kentucky.\\
\textbf{Scenario Results:} Had no issues following the system. I participated to complete the mine unit because only two other testers were present for this session and three drivers were needed.\\
\textbf{Key Feedback:} "I found the system intuitive and easy to navigate."

\begin{table}[ht]
\centering
\caption{Pre-Test Questionnaire Results: Part B - Baseline Safety Knowledge Assessment (n=6). Scale: 1 = Strongly Disagree, 5 = Strongly Agree.}
\label{tab:pretest-safety}
\begin{tabular}{lcccccccc}
\toprule
\textbf{Statement} & \textbf{T1} & \textbf{T2} & \textbf{T3} & \textbf{T4} & \textbf{T5} & \textbf{T6} & \textbf{Mean} & \textbf{SD} \\
\midrule
Q1: Understand primary safety hazards & 5 & 5 & 5 & 5 & 5 & 4 & 4.83 & 0.41 \\
Q2: Familiar with routing procedures & 5 & 5 & 5 & 5 & 5 & 3 & 4.67 & 0.82 \\
Q3: Know how to avoid collisions & 5 & 5 & 5 & 5 & 5 & 4 & 4.83 & 0.41 \\
Q4: Understand safe distances from miner & 5 & 5 & 5 & 5 & 5 & 4 & 4.83 & 0.41 \\
Q5: Aware of communication protocols & 5 & 5 & 5 & 5 & 5 & 4 & 4.83 & 0.41 \\
\midrule
\textbf{Average per Tester} & \textbf{5.0} & \textbf{5.0} & \textbf{5.0} & \textbf{5.0} & \textbf{5.0} & \textbf{3.8} & \textbf{4.80} & \textbf{0.49} \\
\bottomrule
\end{tabular}
\end{table}

\begin{table}[ht]
\centering
\caption{Pre-Test Questionnaire Results: Part C - Expectations and Trust in AI (n=6). Scale: 1 = Strongly Disagree, 5 = Strongly Agree.}
\label{tab:pretest-trust}
\begin{tabular}{lcccccccc}
\toprule
\textbf{Statement} & \textbf{T1} & \textbf{T2} & \textbf{T3} & \textbf{T4} & \textbf{T5} & \textbf{T6} & \textbf{Mean} & \textbf{SD} \\
\midrule
Q1: AI can improve mining safety & 4 & 4 & 3 & 4 & 3 & 5 & 3.83 & 0.75 \\
Q2: Would trust AI for navigation & 3 & 4 & 3 & 4 & 3 & 5 & 3.67 & 0.82 \\
Q3: Concerned about collisions & 5 & 5 & 5 & 5 & 5 & 4 & 4.83 & 0.41 \\
Q4: Expect system to be easy to use & 4 & 4 & 3 & 4 & 3 & 5 & 3.83 & 0.75 \\
Q5: Comfortable with safety-critical AI & 3 & 4 & 3 & 4 & 2 & 5 & 3.50 & 1.05 \\
\midrule
\textbf{Average per Tester} & \textbf{3.8} & \textbf{4.2} & \textbf{3.4} & \textbf{4.2} & \textbf{3.2} & \textbf{4.8} & \textbf{3.93} & \textbf{0.58} \\
\bottomrule
\end{tabular}
\end{table}

\begin{table}[ht]
\centering
\caption{Usability Testing Summary: Quantitative Metrics. Definitions for metrics are provided in Appendix~\ref{sec:questionnaires}.}
\label{tab:usability-quant}
\begin{tabular}{lcccccc}
\toprule
\textbf{Tester} & \shortstack{\textbf{Shuttle}\\\textbf{Car}} & \shortstack{\textbf{Tasks}\\\textbf{Completed}} & \shortstack{\textbf{Response}\\\textbf{Time (s)}} & \textbf{Errors} & \shortstack{\textbf{Agreement}\\\textbf{Rate (\%)}} & \shortstack{\textbf{Adjusted}\\\textbf{AR (\%)}} \\
\midrule
T1 & A & 14/14 & 37 & 0 & 100 & 100 \\
T2 & B & 14/14 & 34 & 1 & 88 & 100 \\
T3 & C & 14/14 & 40 & 1 & 88 & 100 \\
T4 & A & 14/14 & 39 & 0 & 100 & 100 \\
T5 & B & 13/14 & 35 & 2 & 75 & 100 \\
T6 & C & 14/14 & 39 & 0 & 100 & 100 \\
\midrule
\textbf{Average} & -- & 13.8/14 & 37.3 & 0.67 & 91.8 & 100 \\
\bottomrule
\end{tabular}

\vspace{0.5em}
\footnotesize
\textit{Note:} Each task is one round trip from feeder to miner and back. Testers completed two round trips per face (14 total). Response time is the average per task, excluding miner movement and post-task waiting. Errors are incorrect route selections or misunderstandings. Agreement Rate is the percentage of tasks where the tester followed the AI's recommended route without deviation; Adjusted AR (Agreement Rate) excludes corrective steps and confirmed mistakes needed to return to the safe route.
\end{table}

\begin{table}[ht]
\centering
\caption{Usability Testing Summary: Questionnaire-Based Rating Metrics. Cognitive Load and Safety Perception scores are based on questionnaires in Appendix~\ref{sec:questionnaires}.}
\label{tab:usability-qual}
\begin{tabular}{lcccc}
\toprule
\textbf{Tester} & \textbf{Shuttle Car} & \textbf{SUS Score} & \textbf{Cognitive Load (1-7)} & \textbf{Safety Perception (1-5)} \\
\midrule
T1 & A & 92 & 2 & 5 \\
T2 & B & 80 & 3 & 4 \\
T3 & C & 78 & 4 & 4 \\
T4 & A & 90 & 2 & 5 \\
T5 & B & 70 & 5 & 3 \\
T6 & C & 85 & 2 & 5 \\
\midrule
\textbf{Average} & -- & 82.5 & 3.0 & 4.3 \\
\bottomrule
\end{tabular}
\end{table}

\subsection*{Selected Interaction Log Excerpts - Session 1}
\label{sec:log-session1}
The full interface log for Session 1 was substantially larger than is practical to reproduce in full. The excerpt below is intentionally truncated.

\begin{verbatim}
[
    {
        "time": "5:44:50 PM",
        "vehicle": "Shuttle C",
        "direction": "forward",
        "suggestedDirection": "forward",
        "needsFeedback": false,
        "minerPosition": {"gridX": 12, "gridY": 3},
        "inferredPath": [[6,8],[6,7],[6,6],[6,5],...,[12,4],[12,3]]
    },
    {
        "time": "5:44:58 PM",
        "vehicle": "Shuttle C",
        "direction": "right",
        "suggestedDirection": "right",
        "needsFeedback": false,
        "feedback": "This is similar to a video game and would be nice to have play sounds",
        "minerPosition": {"gridX": 12, "gridY": 3},
        "inferredPath": [[6,8],[6,7],[6,6],[6,5],...,[12,4],[12,3]]
    },
    {
        "time": "5:45:05 PM",
        "vehicle": "Shuttle A",
        "direction": "right",
        "suggestedDirection": "forward",
        "needsFeedback": true,
        "feedback": "made wrong turn",
        "minerPosition": {"gridX": 12, "gridY": 3},
        "inferredPath": [[6,8],[6,7],[6,6],[6,5],...,[0,4],[0,3]]
    }
]
\end{verbatim}

\section{Usability Testing Guide and Results - Session 2}
\label{app:usability-trial2}
\subsection*{Tester 1 (T1)}
\textbf{Role:} Female shuttle car operator with 2 years of experience in Kentucky.\\
\textbf{Scenario Results:} Completed all tasks in shuttle car A without deviating from the recommended route.\\
\textbf{Key Feedback:} "Providing feedback while driving was very easy. Before, I could not follow the feedback in the log."

\subsection*{Tester 2 (T2)}
\textbf{Role:} Male mechanic with 15 years of experience in Indiana.\\
\textbf{Scenario Results:} Completed all tasks in shuttle car B without deviating from the recommended route.\\
\textbf{Key Feedback:} "Using the system a second time was much easier.  I'm getting used to it. The shuttle car movement is still slow."

\subsection*{Tester 3 (T3)}
\textbf{Role:} Male face boss with 10 years of experience in Kentucky.\\
\textbf{Scenario Results:} Completed all tasks in shuttle car C without deviating from the recommended route.\\
\textbf{Key Feedback:} No response.

\subsection*{Tester 4 (T4)}
\textbf{Role:} Male Shuttle car operator, 12 years of experience in Kentucky.\\
\textbf{Scenario Results:} Completed all tasks in shuttle car A without deviating from the recommended route.\\
\textbf{Key Feedback:} "Adding the sounds made the system easier to use."

\subsection*{Tester 5 (T5)}
\textbf{Role:} Male Miner operator, 5 years of experience in Kentucky.\\
\textbf{Scenario Results:} Completed all tasks in shuttle car B without deviating from the recommended route.\\
\textbf{Key Feedback:} No response.

\subsection*{Tester 6 (T6)}
\textbf{Role:} Male technical user with previous mining experience in Kentucky.\\
\textbf{Scenario Results:} Completed all tasks in shuttle car C with only a minor deviation in the route.\\
\textbf{Key Feedback:} "Adding the sound was a big improvement."

Since the same participants were involved in both usability testing sessions, the pre-test questionnaires from the first session were reused for consistency.

\begin{table}[ht]
\centering
\caption{Usability Testing Summary - Session 2: Quantitative Metrics. Definitions for metrics are provided in Appendix~\ref{sec:questionnaires}.}
\label{tab:usability-quant-trial2}
\begin{tabular}{lcccccc}
\toprule
\textbf{Tester} & \shortstack{\textbf{Shuttle}\\\textbf{Car}} & \shortstack{\textbf{Tasks}\\\textbf{Completed}} & \shortstack{\textbf{Response}\\\textbf{Time (s)}} & \textbf{Errors} & \shortstack{\textbf{Agreement}\\\textbf{Rate (\%)}} & \shortstack{\textbf{Adjusted}\\\textbf{AR (\%)}} \\
\midrule
T1 & A & 14/14 & 31 & 0 & 100 & 100 \\
T2 & B & 14/14 & 32 & 0 & 100 & 100 \\
T3 & C & 14/14 & 34 & 0 & 100 & 100 \\
T4 & A & 14/14 & 32 & 0 & 100 & 100 \\
T5 & B & 14/14 & 35 & 0 & 100 & 100 \\
T6 & C & 14/14 & 33 & 1 & 96 & 100 \\
\midrule
\textbf{Average} & -- & 14.0/14 & 32.8 & 0.17 & 99.3 & 100 \\
\bottomrule
\end{tabular}

\vspace{0.5em}
\footnotesize
\textit{Note:} Each task is one round trip from feeder to miner and back. Testers completed two round trips per face (14 total). Response time is the average per task, excluding miner movement and post-task waiting. Errors are incorrect route selections or misunderstandings. Agreement Rate is the percentage of tasks where the tester followed the AI's recommended route without deviation; Adjusted AR (Agreement Rate) excludes corrective steps and confirmed mistakes needed to return to the safe route.
\end{table}

\begin{table}[ht]
\centering
\caption{Usability Testing Summary - Session 2: Questionnaire-Based Rating Metrics. Cognitive Load and Safety Perception scores are based on questionnaires in Appendix~\ref{sec:questionnaires}.}
\label{tab:usability-qual-trial2}
\begin{tabular}{lcccc}
\toprule
\textbf{Tester} & \textbf{Shuttle Car} & \textbf{SUS Score} & \textbf{Cognitive Load (1-7)} & \textbf{Safety Perception (1-5)} \\
\midrule
T1 & A & 95 & 1 & 5 \\
T2 & B & 88 & 2 & 5 \\
T3 & C & 84 & 3 & 4 \\
T4 & A & 94 & 1 & 5 \\
T5 & B & 82 & 3 & 4 \\
T6 & C & 90 & 2 & 5 \\
\midrule
\textbf{Average} & -- & 88.8 & 2.0 & 4.7 \\
\bottomrule
\end{tabular}
\end{table}

\subsection*{Selected Interaction Log Excerpts - Session 2}
\label{sec:log-session2}
As with Session 1, the full Session 2 log is truncated.

\begin{verbatim}
[
    {
        "time": "6:45:05 PM",
        "vehicle": "Shuttle A",
        "direction": "right",
        "suggestedDirection": "right",
        "needsFeedback": false,
        "minerPosition": {"gridX": 10, "gridY": 3},
        "inferredPath": [[6,8],[7,8],[8,8],[8,7],...,[10,4],[10,3]]
    },
    {
        "time": "6:45:09 PM",
        "vehicle": "Shuttle B",
        "direction": "forward",
        "suggestedDirection": "forward",
        "needsFeedback": false,
        "minerPosition": {"gridX": 10, "gridY": 3},
        "inferredPath": [[6,8],[6,7],[6,6],[7,6],...,[10,4],[10,3]]
    },
    {
        "time": "6:45:34 PM",
        "vehicle": "Shuttle C",
        "direction": "left",
        "suggestedDirection": "forward",
        "needsFeedback": true,
        "feedback": "negative test",
        "minerPosition": {"gridX": 8, "gridY": 3},
        "inferredPath": [[6,8],[6,7],[6,6],[6,5],...,[8,4],[8,3]]
    },
    {
        "time": "6:45:43 PM",
        "vehicle": "Shuttle C",
        "direction": "forward",
        "suggestedDirection": "forward",
        "needsFeedback": false,
        "minerPosition": {"gridX": 8, "gridY": 3},
        "inferredPath": [[6,8],[6,7],[6,6],[6,5],...,[8,4],[8,3]]
    }
]
\end{verbatim}

\section{Questionnaires}
\label{sec:questionnaires}

\subsection{Pre-Test Materials}

\subsubsection{Pre-Test Questionnaire}
\label{sec:pretest-questionnaire}

\textbf{Part A: Background Information}

\begin{enumerate}[leftmargin=*]
    \item What is your age range?
    \begin{itemize}
        \item[$\square$] 18-25 \quad $\square$ 26-35 \quad $\square$ 36-45 \quad $\square$ 46-55 \quad $\square$ 56+
    \end{itemize}

    \item What is your gender?
    \begin{itemize}
        \item[$\square$] Male \quad $\square$ Female \quad $\square$ Non-binary \quad $\square$ Prefer not to say
    \end{itemize}

    \item What state or geographic region do you work in?
    \begin{itemize}
        \item[$\square$] Kentucky \quad $\square$ West Virginia \quad $\square$ Pennsylvania \quad $\square$ Indiana \quad $\square$ Other: \_\_\_\_\_\_\_\_\_\_\_
    \end{itemize}

    \item Do you have prior experience working in underground coal mining?
    \begin{itemize}
        \item[$\square$] Yes (If yes, how many years? \_\_\_\_\_) \quad $\square$ No
    \end{itemize}

    \item Have you ever operated a shuttle car in a mine?
    \begin{itemize}
        \item[$\square$] Yes (If yes, how many years? \_\_\_\_\_) \quad $\square$ No
    \end{itemize}

    \item What is your level of experience with navigation or routing systems? (e.g., GPS, Google Maps)
    \begin{itemize}
        \item[$\square$] None \quad $\square$ Basic \quad $\square$ Intermediate \quad $\square$ Advanced
    \end{itemize}

    \item What is your level of comfort with technology in general?
    \begin{itemize}
        \item[$\square$] Very uncomfortable \quad $\square$ Somewhat uncomfortable \quad $\square$ Neutral \quad $\square$ Somewhat comfortable \quad $\square$ Very comfortable
    \end{itemize}
\end{enumerate}

\textbf{Part B: Baseline Safety Knowledge Assessment}

Rate your agreement with the following statements on a scale of 1--5 (1 = Strongly Disagree, 5 = Strongly Agree):

\begin{enumerate}[leftmargin=*]
    \item I understand the primary safety hazards in underground coal mining operations.
    \begin{itemize}
        \item[$\square$] 1 \quad $\square$ 2 \quad $\square$ 3 \quad $\square$ 4 \quad $\square$ 5
    \end{itemize}

    \item I am familiar with standard shuttle car routing procedures in room-and-pillar mines.
    \begin{itemize}
        \item[$\square$] 1 \quad $\square$ 2 \quad $\square$ 3 \quad $\square$ 4 \quad $\square$ 5
    \end{itemize}

    \item I know how to avoid collisions when multiple shuttle cars operate simultaneously.
    \begin{itemize}
        \item[$\square$] 1 \quad $\square$ 2 \quad $\square$ 3 \quad $\square$ 4 \quad $\square$ 5
    \end{itemize}

    \item I understand the importance of maintaining safe distances from the continuous miner.
    \begin{itemize}
        \item[$\square$] 1 \quad $\square$ 2 \quad $\square$ 3 \quad $\square$ 4 \quad $\square$ 5
    \end{itemize}

    \item I am aware of communication protocols used between shuttle car operators and miners.
    \begin{itemize}
        \item[$\square$] 1 \quad $\square$ 2 \quad $\square$ 3 \quad $\square$ 4 \quad $\square$ 5
    \end{itemize}
\end{enumerate}

\textbf{Part C: Expectations and Trust in AI}

Rate your agreement with the following statements on a scale of 1--5 (1 = Strongly Disagree, 5 = Strongly Agree):

\begin{enumerate}[leftmargin=*]
    \item I believe AI systems can improve safety in underground mining operations.
    \begin{itemize}
        \item[$\square$] 1 \quad $\square$ 2 \quad $\square$ 3 \quad $\square$ 4 \quad $\square$ 5
    \end{itemize}

    \item I would trust an AI system to provide navigation recommendations in a mine.
    \begin{itemize}
        \item[$\square$] 1 \quad $\square$ 2 \quad $\square$ 3 \quad $\square$ 4 \quad $\square$ 5
    \end{itemize}

    \item I am concerned about collisions or accidents when multiple shuttle cars operate simultaneously.
    \begin{itemize}
        \item[$\square$] 1 \quad $\square$ 2 \quad $\square$ 3 \quad $\square$ 4 \quad $\square$ 5
    \end{itemize}

    \item I expect the shuttle car navigation system to be easy to use.
    \begin{itemize}
        \item[$\square$] 1 \quad $\square$ 2 \quad $\square$ 3 \quad $\square$ 4 \quad $\square$ 5
    \end{itemize}

    \item I am comfortable using computer-based systems to assist with safety-critical tasks.
    \begin{itemize}
        \item[$\square$] 1 \quad $\square$ 2 \quad $\square$ 3 \quad $\square$ 4 \quad $\square$ 5
    \end{itemize}
\end{enumerate}

\subsection{Cognitive Load Questionnaire}
\label{sec:cogload-questionnaire}
Participants rated each item on a 1--7 scale (1 = Very Low, 7 = Very High):

\begin{enumerate}[leftmargin=*]
    \item How mentally demanding was it to operate the shuttle car navigation system?
    \item How much effort did you have to invest to complete tasks using the system?
    \item How difficult was it to keep track of your location and route recommendations?
    \item How much concentration was required to use the system effectively?
    \item How often did you feel overloaded with information from the navigation system?
    \item How easy was it to understand the instructions provided by the system?
    \item How challenging was it to make decisions based on the system's recommendations?
    \item How much did you have to remember while using the navigation system?
    \item How stressful was it to use the system during mining operations?
    \item How confident were you in your ability to use the system without making mistakes?
\end{enumerate}

\subsection{Safety Perception Questionnaire}
\label{sec:safety-questionnaire}
Participants rated each item on a 1--5 scale (1 = Strongly Disagree, 5 = Strongly Agree):

\begin{enumerate}[leftmargin=*]
    \item How safe would you feel using the shuttle car navigation system underground, if it were available?
    \item Did the system help you avoid hazardous areas or situations?
    \item Did the navigation recommendations increase your confidence in safe route selection?
    \item How well did the system alert you to potential risks or dangers?
    \item Did you trust the system to prioritize your safety during mining operations?
    \item Did the system reduce your concerns about collisions or accidents?
    \item How effective was the system in supporting safe shuttle car operation?
    \item Did the system provide clear information about safety-related decisions?
    \item Did you feel the system improved overall safety in the mine?
    \item Would you recommend the system to other operators for safer shuttle car navigation?
\end{enumerate}

\section{SHAP Analysis}
\label{sec:shap-analysis}
This section summarizes the post-evaluation SHAP analysis for the validated deployed model. For these scenario-level features, \texttt{dest\_x} represents the horizontal grid position and \texttt{dest\_y} represents the vertical grid position of the destination. In the evaluated mining scenarios, the miner remains within the same cross cut, so \texttt{dest\_y} never changes and therefore has little importance for the model. By contrast, \texttt{dest\_x} captures the miner's movement across the face and is the most important feature in determining routes. The second most important factor is shuttle-car identity, reflected in features such as \texttt{is\_car\_A}, because each shuttle car has its own predetermined safe route relative to the miner. It is also important to note that the AI model determines only the path from the feeder to the miner. The user interface is responsible for managing direction of travel, including movement from the feeder to the miner and the return trip from the miner back to the feeder. The tables below report the accuracy check performed after the SHAP workflow, the mean absolute SHAP values for the validated start-of-route decision, and grouped parameter importance across the scenario-level features.

\begin{table}[ht]
\centering
\caption{Post-evaluation accuracy check for SHAP}
\label{tab:shap-accuracy-check}
\begin{tabular}{lccc}
\toprule
\textbf{Destination} & \textbf{A} & \textbf{B} & \textbf{C} \\
\midrule
(0, 3) & 1.0 & 1.0 & 1.0 \\
(2, 3) & 1.0 & 1.0 & 1.0 \\
(4, 3) & 1.0 & 1.0 & 1.0 \\
(6, 3) & 1.0 & 1.0 & 1.0 \\
(8, 3) & 1.0 & 1.0 & 1.0 \\
(10, 3) & 1.0 & 1.0 & 1.0 \\
(12, 3) & 1.0 & 1.0 & 1.0 \\
\bottomrule
\end{tabular}
\end{table}

\begin{table}[ht]
\centering
\caption{Mean absolute SHAP values for the validated start-of-route decision}
\label{tab:shap-feature-importance}
\begin{tabular}{clc}
\toprule
\textbf{Rank} & \textbf{Feature} & \textbf{Mean Absolute SHAP} \\
\midrule
0 & \texttt{dest\_x\_norm} & $1.176640 \times 10^{-6}$ \\
1 & \texttt{is\_car\_C} & $5.679787 \times 10^{-7}$ \\
2 & \texttt{is\_car\_B} & $5.272961 \times 10^{-7}$ \\
3 & \texttt{is\_car\_A} & $1.658838 \times 10^{-7}$ \\
4 & \texttt{dest\_y\_norm} & $0.0$ \\
\bottomrule
\end{tabular}
\end{table}

\begin{table}[ht]
\centering
\caption{Grouped parameter importance from SHAP analysis}
\label{tab:shap-group-importance}
\begin{tabular}{lc}
\toprule
\textbf{Parameter Group} & \textbf{Mean Absolute SHAP} \\
\midrule
\texttt{shuttle\_identity} & $1.0 \times 10^{-6}$ \\
\texttt{destination\_x} & $1.0 \times 10^{-6}$ \\
\texttt{destination\_y} & $0.0$ \\
\bottomrule
\end{tabular}
\end{table}

\end{document}